\documentclass[a4paper,fleqn,usenatbib,twocolumn,letters]{mnras}

\usepackage[T1]{fontenc}
\usepackage{ae,aecompl}

\usepackage{graphicx}	% Including figure files
\usepackage{amsmath}	% Advanced maths commands
\usepackage{amssymb}	% Extra maths symbols
\usepackage{bm}
\usepackage{xspace}

\newcommand{\ac}{Astron. \& Comput.\xspace}

\title[WASP-4 transit timing variation]{WASP-4 transit timing variation from a comprehensive set of 129 transits}

\author[Baluev et~al.]{%
R.V.~Baluev$^{1,2}$\thanks{E-mail: r.baluev@spbu.ru}, E.N.~Sokov$^{1,2}$, S.~Hoyer$^3$, C.~Huitson$^4$, Jos\'e A.R.S. da~Silva$^5$,
\newauthor P.~Evans$^6$, I.A.~Sokova$^{1,2}$, C.R.~Knight$^7$, V.Sh.~Shaidulin$^1$
\\
$^1$Saint Petersburg State University, Universitetskaya emb. 7--9, St Petersburg 199034, Russia\\
$^2$Central Astronomical Observatory at Pulkovo of Russian Academy of Sciences, Pulkovskoje sh. 65/1, St Petersburg 196140, Russia\\
$^3$Aix Marseille Univ, CNRS, CNES, LAM, Marseille, France\\
$^4$CASA, University of Colorado, 389 UCB, Boulder, CO, 80309-0389, USA\\
$^5$Department of Mathematical Sciences, University of South Africa, Private Bag X6, Florida 1709, South Africa\\
$^6$El Sauce Observatory, Coquimbo Province, Chile\\
$^7$Ngileah Observatory, 144 Kilkern Road, RD 1. Bulls 4894, New Zealand\\
}

\date{Accepted 2020 April 20.
      Received 2020 April 17;
      in original form 2020 March 10}

\pubyear{2020}

\begin{document}
\label{firstpage}
\pagerange{\pageref{firstpage}--\pageref{lastpage}}
\maketitle

\begin{abstract}
We homogeneously reanalyse $124$ transit light curves for the WASP-4~b hot Jupiter. This
set involved new observations secured in 2019 and nearly all observations mentioned in the
literature, including high-accuracy GEMINI/GMOS transmission spectroscopy of 2011--2014 and
TESS observations of 2018. The analysis confirmed a nonlinear TTV trend with $P/|\dot
P|\sim (17-30)$~Myr (1-sigma range), implying only half of the initial decay rate
estimation. The trend significance is at least $3.4$-sigma in the agressively conservative
treatment. Possible radial acceleration due to unseen companions is not revealed in Doppler
data covering seven years 2007--2014, and radial acceleration of $-15$~m~s$^{-1}$yr$^{-1}$
reported in a recent preprint by another team is not confirmed. If present, it is a very
nonlinear RV variation. Assuming that the entire TTV is tidal in nature, the tidal quality
factor $Q_\star'\sim (4.5-8.5)\cdot 10^4$ does not reveal a convincing disagreement with
available theory predictions.
\end{abstract}


%%%%%%%%%%%%%%%%%%%%%%%%%%%%%%%%%%%%%%%%%%%%%%%%%%

%%%%%%%%%%%%%%%%% BODY OF PAPER %%%%%%%%%%%%%%%%%%

\section{Introduction}
The nonlinear trend in WASP-4~b transit timings was initially claimed by \citet{Bouma19},
based on new TESS observations acquired in 2018, and on ground-based observations published
so far. According to their fit, its magnitude (curvature) corresponded to the period decay
parameter $P/|\dot P| \sim 9.2$~Myr. One explanation of such a TTV trend is planet-star
tidal interaction, causing either (i) planet spiraling on to the star, or (ii) tidal
apsidal drift of a weakly eccentric planetary orbit. If the TTV trend and its tidal nature
is confirmed, this target would be the second such example known, in addition to WASP-12.

However, in \citep{Baluev19} some doubts were raised concerning the reality of this TTV
trend for WASP-4. In particular, it appeared that the trend is sensitive to just a few
transit timings that may be affected by red noise or other systematic effects. Also,
spurious TTV effects might appear because of the heterogeneous nature of the timings. Since
different transit observations were processed by different teams using different models and
approaches, they may have different systematic errors. Even more importantly, timing
uncertainties coming from different teams may have systematically different biases. Then
some transit times with overstated accuracy may inadequately dominate over others,
corrupting the results.

The TTV trend was neither confirmed nor retracted in \citep{Baluev19}, but soon after that
\citet{Southworth19} presented a set of $22$ additional WASP-4 transit observations
apparently confirming its TTV trend at a high significance. However, they obtain a smaller
TTV curvature than \citet{Bouma19}, while their analysis still did not take into account
the issue of inhomogeneous data. Simultaneously, \citet{Baluev19} did not reprocess two
crucial sets of transit lightcurves from \citep{Hoyer13,Huitson17}, because they were not
published in table form. Without them, that homogeneous analysis of WASP-4 was not
complete.

Therefore, TTV studies currently available for WASP-4 are still disputable and incomplete
in one concern or another. In this work our goal is to homogeneously analyse the
comprehensive set of nearly all ground-based WASP-4 transit observations, including those
from \citep{Hoyer13,Huitson17,Southworth19}, and to verify the existence of the TTV trend.

In Sect.~\ref{sec_src} we describe photometric data and their processing, while in
Sect.~\ref{sec_res} we present the main analysis results.

\section{Input data and their processing}
\label{sec_src}
\subsection{Transit photometry}
We used most of the lightcurves from \citep{Baluev19}. This includes
\citep{Wilson08,Gillon09a,Sanchis-Ojeda11,Nikolov12,Petrucci13}, and amateur observations.

Recently, \citep{Southworth19} published a large amount of WASP-4 transit observations,
including a replacement for \citep{Southworth09b}, as well as new light curves. Previously
there were notices about clock errors affecting the Danish telescope used by
\citet{Southworth09b}, motivating \citet{Baluev15a,Baluev19} to remove those data from
the TTV analysis. However, according to \citet{Southworth19} that errors did not affect any
WASP-4 observations. Therefore, in this work we use all these data without restriction.

We now use previously unpublished transit lightcurves from \citep{Hoyer13} and from
\citep{Huitson17}. The latter data are especially important, because they are of a very
high accuracy, derived from the transmission spectroscopy at GEMINI. However, it was
noticed in \citep{Baluev19} that transit times from \citep{Huitson17} might have
understated uncertainties, possibly resulting in spuriuous or biased TTVs.

We also reanalysed $18$ public TESS lightcurves available in the MAST (Mikulski archive).
We removed all data that had nonzero quality indicator and cut out pieces $\pm 0.2$~d
around each transit (about $4$ transit durations). The rest of the light curve was not used
as it showed obvious hints of weak variability.

In 2019 we also obtained $6$ new EXPANSION amateur observations of WASP-4, which are also
included in this compilation. We therefore have now $124$ transit lightcurves to be
reprocessed homogeneously, compared to only $66$ ones reprocessed by \citet{Baluev19}.

Finally, we used $5$ old timings found in \citep{Gillon09a,Dragomir11} without lightcurves,
increasing the total transits number to $N=129$. These timings are not so accurate, and
their effect remains small.

As in \citep{Baluev19}, we do not include the HST observations by \citet{Ranjan14} because
they were partly overexposed.

\subsection{Lightcurves fitting and analysis}
We used entirely the same processing pipeline as in \citep{Baluev19}, simply adding new
lightcurves at Stage~4 (full dataset) and Stage~5 (HQ dataset). The only major difference
is that now we determine the HQ subset in a different more useful manner (detailed below).

Some important details of the algorithm are as follows. The lightcurves are fitted
simultaneously against $3$ planetary parameters common for all transits (the planet/star
radii ratio, orbital impact parameter, transit duration), and individual transit timings.
We also fit cubic trend in each lightcurve, and quadratic limb-darkening law on a
band-per-band basis. The fitting is done using {\sc PlanetPack3} software \citep{Baluev18c}
through an advanced maximum-likelihood approach with regularized noise model described in
\citep{Baluev08b,Baluev15a}.

The resulting TTV data were fitted by a quadratic trend model with a curvature parameter
$q=-\dot P/P$. This TTV model is linear with respect to its parameters, so even if $q$ is
small or zero, its uncertainty range remains meaningful and is nearly symmetric. The
inverse $T_{\rm d}=q^{-1}$ is perhaps a more intuitive quantity, the so-called period decay
timescale, but it is not a linear parameter. If the uncertainty of $\sigma_q$ appears
comparable to $q$ then confidence ranges for $T_{\rm d}$ are not symmetric.

This fitting was done via the maximum likelihood approach involving fittable noise
\citep{Baluev08b}, and using either multiplicative or so-called regularized noise model
\citep{Baluev14a}. That regularized noise model behaves mostly similar to the one with an
additive ``jitter'' \citep{Wright05}, but mitigates some of its singular behaviour. For the
full $N=129$ dataset, we considered a total of $4$ distinct models of the TTV noise, also
depending on whether we treat this dataset as a whole or put the spotted transits (see
below) into a separate subset with an independent noise model. See \citep{Baluev19} for
further discussion of possible TTV noise models and their role.

\subsection{Filtering out potentially currupted lightcurves}
WASP-4 demonstrates obvious spotting activity, and starspots appear as anomalies in the
transit lightcurves \citep{Sanchis-Ojeda11}. Such events disturb the shape of the
lightcurve, building up additional noise in the derived transit times. According to
\citet{Baluev19}, those errors may increase the TTV residuals r.m.s. by $\sim 30-100$ per
cent, and may reach $\sim 2$~min (the case of HD~189733); see also
\citep{Barros13,Oshagh13}. The putative WASP-4 TTV is below $1$~min, so it can be
significantly distorted by such errors.

This motivated us to apply some filter removing potentially unreliable transits, affected
by spots. Notice that \citet{Southworth19} undertook a similar work, but only for their own
portion of data. Now we try to do this for the entire set of $124$ lightcurves we have.

The first filter was targeted to remove transits affected by spots. For each lightcurve we
computed the residuals r.m.s. $s_{\rm in/out}$ for the in-transit and out-of-transit
portions. If their ratio $\rho = s_{\rm in}/s_{\rm out}$ is large (compared to its
uncertainty $\sigma_\rho$) then we may deal with a spot-transit anomaly. However, we
avoided any formal thresholds on $\rho$, because this quantity is merely a rough indicator,
e.g. it does not take into account any red noise. Instead, we adopted a simplified
approach. We formed two ranged lists: one with increasing $\rho$ and the other one with
increasing significance of $\log\rho$, or $(\rho \log\rho)/\sigma_\rho$. After that, we
extracted top quartiles from the both lists. It appeared that both these top quartiles
contained almost the same transits (with only a few exceptions). We finally constructed
their intersection, forming the list of unreliable transits with only \emph{significantly}
large $\rho$. Finally, we also removed a few lightcurves that did not necessarily reveal
clear spotting activity themselves, but were made at one of the same dates.

The second filter is a mixture of a periodogram analysis and jacknife. First, we noticed
that the periodogram of TTV residuals (for our full dataset) demonstrated multiple spurious
peaks that appear tall enough to pass formal significance threshold but visually looked
like noise (similar to e.g. \citealt{Baluev15a}). Since this behaviour appears quite usual
for WASP-4 and for a few other spotted targets, we suspected that the periodogram may
simply reveal sensitivity to one or a few biased timings (corrupted by spots or other
systematic errors). So to detect such bad points we subsequently removed one of them and
recomputed the periodogram for each such reduced dataset. If the maximum periodogram peak
changed too much after that, such a transit was identified as unreliable. The exact
thresholds were subjective here, as we only wanted to identify clearly odd transits or
those that offer abnormally high contribution into the results. We found just a few such
odd lightcurves. After that, we also computed the formal significance of the quadratic TTV
trend as $q/\sigma_q$ and processed it in the similar way, identifying timings that
contribute too remarkably. But this eventually did not highlight any new transits.

In the end of this procedure we had $89$ HQ lightcurves out of the initial $124$ ones.
These two sets were passed through our primary pipeline independently. Thus, together with
$5$ third-party timings, we constructed two TTV datasets: the full one ($129$ points) and a
homogeneous HQ one ($89$ points). They were obtained independently from each other, and are
attached as online-only material (the format of the files is the same as in
\citealt{Baluev15a}).

\begin{table}
\caption{WASP-4: summary of transit data}
\label{tab_data}
\begin{tabular}{lcc}
\hline
Publication/Team        & transits & unreliable \\
\hline
\citet{Sanchis-Ojeda11} &  $6$ & $4$ \\
\citet{Nikolov12}       & $12$ & $4$ \\
\citet{Petrucci13}      &  $6$ & $2$ \\
\citet{Hoyer13}         & $12$ & $2$ \\
\citet{Huitson17}       &  $4$ & $3$ \\
\citet{Southworth19}    & $22$ &$10$ \\
TRAPPIST                &  $6$ & $1$ \\
TESS-2018               & $18$ & $3$ \\
others (mostly amateurs)& $38$ & $6$ \\
\hline
Total                   &$124$ & $35$\\
\hline
\end{tabular}
\end{table}

A summary of the filtered data is given in Table~\ref{tab_data}. As we can see, the most
accurate observations reveal large fraction of potentially corrupted lightcurves, e.g.
$3/4$ of \citep{Huitson17}, $2/3$ of \citep{Sanchis-Ojeda11}, and $1/2$ of
\citep{Southworth19}, though only $1/6$ of TESS-2018. On contrary, most amateur
observations did not reveal detectable spots. Although expected, such behaviour is rather
regretful, because the most accurate timings should likely have understated uncertainties
and are not reliable because of that.

Notice that our filtering procedure was not supposed to be comprehensive, and several
lightcurves in our HQ sample still demonstrate clear spot anomalies. For example, our
method revealed $5$ of $6$ spots-affected lightcurves noticed by \citet{Southworth19},
additionally blacklisting $5$ other their lightcurves. \citet{Hoyer13} identified $4$
possibly spotted transits, but none coinciding with our two. Simultaneously, they did not
detect a spot in the transit of 2008/10/01 independently observed by \citet{Southworth19}.
Given such inconsistencies, we assume it might be difficult to distinguish spot-transit
anomalies from various systematic errors. Our goal here was to remove a predefined moderate
fraction of empirically most unreliable lightcurves and to see how much the results would
change then.

\subsection{Doppler data and a self-consistent analysis}
Any TTV trend may be an apparent effect of the unseen companion that causes variable light
arrival delay (the Roemer effect, \citealt{Irwin52}). If the period of the companion is
larger than the observation time span, it may induce a piece of a sinusoidal TTV variation
following close to a parabola. However, such a companion would also reveal itself through
radial acceleration (RA) of the host star. Therefore, Doppler measurement may help to
discard such interpretation, or to put useful constraints.

We used here $99$ radial velocity (RV) measurements available in \citep{Baluev19}. This
includes $49$ HARPS measurements derived with HARPS-TERRA \citep{AngladaEscudeButler12},
$45$ CORALIE ones derived with a standard CORALIE reduction pipeline, and $5$ Keck
measuments from \citep{Knutson14}.

The self-consistent fitting of the transit lightcurves and RV data was performed using the
{\sc PlanetPack} software \citep{Baluev13c,Baluev18c}. The details are entirely the same as
in \citep{Baluev19}, in particular we adopted a fixed star mass of $0.930$~$M_\odot$ from
\citep{Triaud10}.

\section{Results}
\label{sec_res}
Our full TTV dataset, including all $129$ transits, is shown in Fig.~1 of the online
supplement, compared with some previous \citep{Bouma19,Southworth19} and new fits of the
TTV trend. Significant deviations are revealed between different trend models. According to
\citet{Southworth19}, their estimate differs from \citet{Bouma19} by about 2-sigma. This
is rather large and possibly indicates some hidden systematic errors.

Our full TTV dataset suggests that the trend magnitude is now decreased even further, to
about half of the value initially claimed \citep{Bouma19}. It still depends on the noise
model, but corresponds to $T_{\rm d}\sim 15-20$~Myr, compared to $9.2$~Myr following from
\citep{Bouma19}. The trend significance appears large, about $5$--$6$ sigma, also
model-dependent.

To estimate the trend significance we followed the same approach as in \citep{Baluev19}.
Given some probe value of $q$, we may compute the log likelihood-ratio statistic $Z$
corresponding to the models with the best fitting $\hat q$ and with the given $q$. The
graph of $Z(q)$ should then have a nearly parabolic shape centered at $\hat q$. The value
$Z(0)$ characterizes the significance of the TTV trend itself (how much $q=0$ agrees with
the data), while horizontal $Z$-levels would determine confidence ranges for $q$, through
intersections with $Z(q)$. As in \citep{Baluev19}, we consider several significance levels
from the $\chi^2$ test, and the BIC ones.

\begin{figure*}
\includegraphics[width=0.49\linewidth]{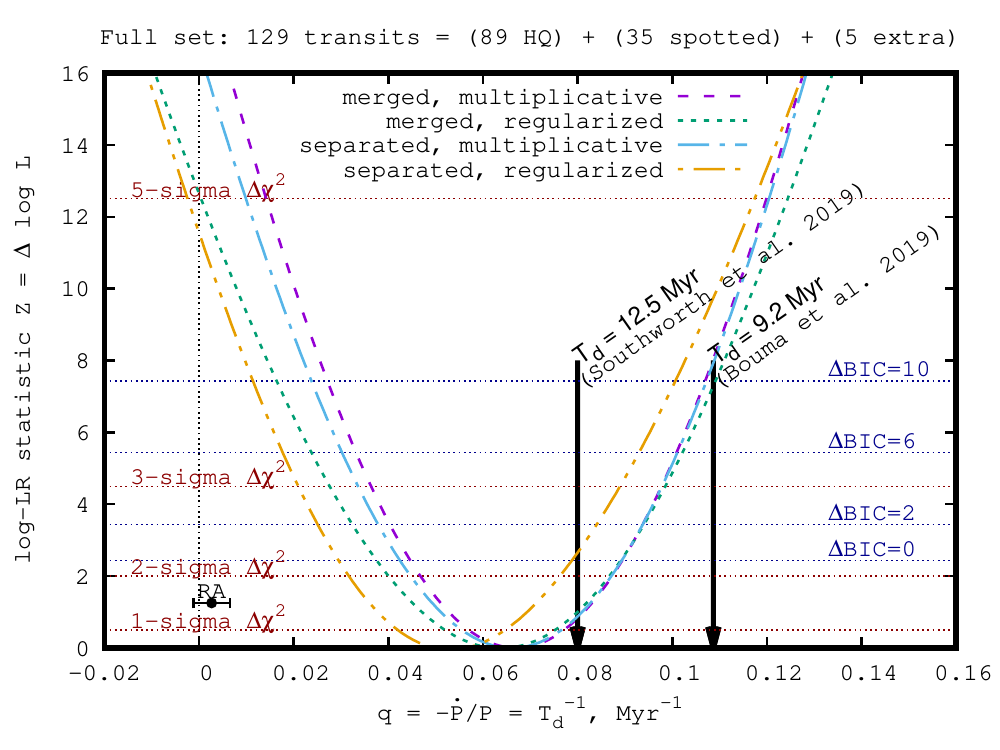}
\includegraphics[width=0.49\linewidth]{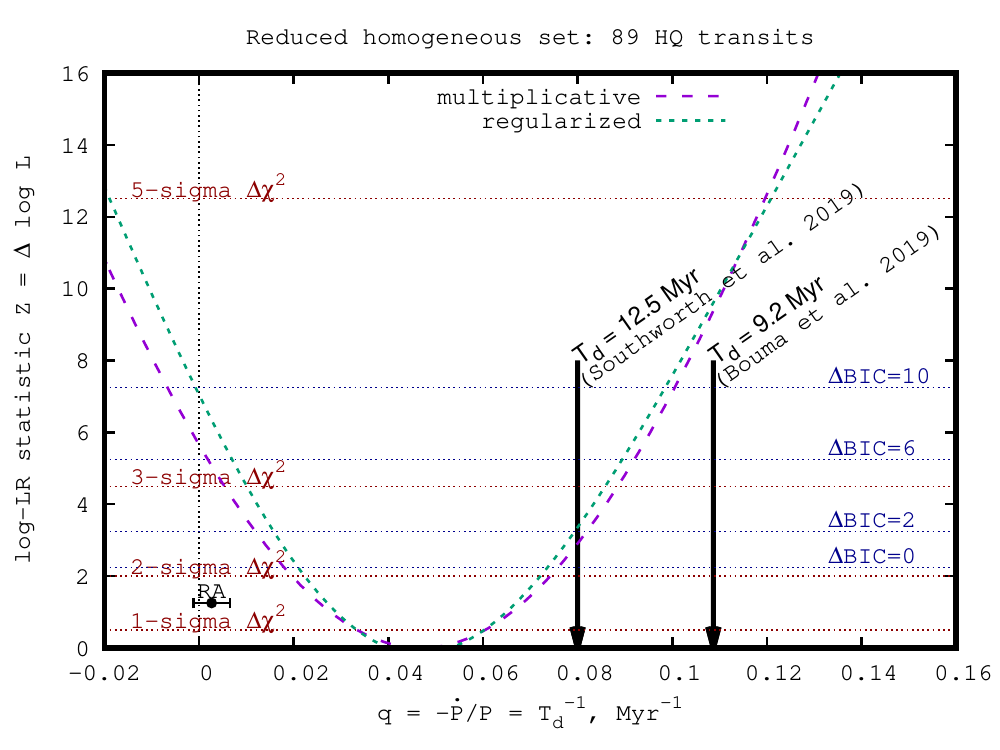}\\
\caption{Logarithm of the likelihood ratio statistic for WASP-4, $Z(q)$, as a function of
$q = -\dot P/P$. Left panel is for the full dataset, right one is for the HQ one.
Near-parabolic curves within each graph correspond to different models of the TTV noise, as
labelled. In each graph a set of the significance threshold levels is shown, corresponding
to the frequentist $\chi^2$ test or to the BIC. The Roemer effect due to possible radial
acceleration (RA) is shown as a horizontal errorbar. See text for more details.}
\label{fig_WASP4-lik}
\end{figure*}

Such plots are shown in Fig.~\ref{fig_WASP4-lik}. For the full TTV dataset $Z(0)$ becomes
very high, above $5$ sigma, but the shape of $Z(q)$ is sensitive to the noise model. Those
model-variable changes correspond to shifts of about $1$-sigma in terms of $q$. They might
indicate that our full TTV dataset is still imperfect statistically. Some of its timings
may be affected by systematic errors, e.g. caused by spot-transit events.

For the HQ subset, the trend significance is reduced to $3.4$--$3.8$ sigma. The best
quadratic fit for the TTV trend then becomes
\begin{eqnarray}
\text{TTV}(n) = T_0 + n P + q \frac{(nP)^2}{2}, \nonumber\\
T_0 = (2455937.080285 \pm 3.9\cdot 10^{-5})\ \text{BJD}_\text{TDB}, \nonumber\\
P = (1.338231472 \pm 3.0\cdot 10^{-8})\ \text{d}, \nonumber\\
q = (47 \pm 13)\ \text{Gyr}^{-1}
\label{ttv}
\end{eqnarray}
It is now almost model-invariable. This value of $q$ corresponds to $T_{\rm d}\simeq
21$~Myr and is close to the smallest value derived from the full TTV dataset. Therefore,
this trend remains significant enough even in our agressively conservative treatment, but
its magnitude is smaller than previously claimed.

More importantly, $Z(q)$ became now considerably less model-sensitive. This might confirm
that (i) we generally correctly removed the most unreliable lightcurves from our analysis,
and (ii) the results derived from the HQ subset are statistically robust. Therefore,
further removal of spotted lightcurves does not look very sensible. This would decrease the
trend significance arbitrarily further, but simply because of reducing the number of
available timings.

We can see that the full data ($129$ transits) already implied a reduction of $q$
relatively to \citet{Southworth19}. This reduction was mainly because old data were
reprocessed and old timings were corrected (not because of new data of 2019 that did not
appear very accurate). However, after rejecting unreliable transits, $q$ is reduced yet
more. Therefore, these two effects join roughly 50/50.

\begin{table}
\caption{WASP-4 self-consistent fit of transit and RV data.}
\label{tab_fit}
\begin{tabular}{lc}
\hline
num. of lightcurves $N_{\rm LC}$        & $89$ (HQ homogeneous)\\
num. of RVs $N_{\rm RV}$                & $99$ \\
\hline
star mass $M_\star$ [$M_\odot$]         & $0.93$ (fixed)\\
star radius $R_\star$ [$R_\odot$]       & $0.9019(47)$\\
star density $\rho_\star$ [$\rho_\odot$]& $1.268(20)$\\
rotation vel. $v\sin i$ [m s$^{-1}$]    & $1970(260)$ \\
spin-orbit ang. $\lambda$ [$^\circ$]    & $341(19)$\\
rad. accel. $c_1$ [m~s$^{-1}$~yr$^{-1}$]& $-0.8(1.2)$\\
\hline
planet mass $m$ [$M_{\rm J}$]           & $1.1974(68)$ \\
planet radius $r$ [$R_{\rm J}$]         & $1.3846(88)$ \\
orbital period$^{1,2}$ $P$ [d]          & $1.338231602(56)$\\
TTV trend$^2$ $T_{\rm d}$ [Myr]         & $21.3(5.5)$ \\
mean longitude$^1$ $l$ [$^\circ$]       & $235.86(26)$ \\
inclination $i$ [$^\circ$]              & $88.87(39)$ \\
eccentricity $e$                        & $0.0053(38)$ \\
pericenter arg. $\omega$ [$^\circ$]     & $247(28)$ \\
$e\cos\omega$                           & $-0.0021(22)$ \\
$e\sin\omega$                           & $-0.0049(41)$ \\
\hline
\end{tabular}
\begin{flushleft}
\tiny
The fitting uncertainties are given in parenthesis after each estimation, in the units of
the last two figures. The $M_\star$ uncertainty was not included in the fit.\\ $^1$These
parameters refer to $T_0=2455197.5$ (2015/01/01) in the BJD TDB system.\\ $^2$The
uncertainty in $P$ and $T_{\rm d}$ should be scaled up by the TTV scatter $\sqrt{\chi_{\rm
TTV}^2}=1.14$ to include the effect of remaining spots. Notice that $P$ here is different
from~(\ref{ttv}) because they refer to different epochs.
\end{flushleft}
\end{table}

In Table~\ref{tab_fit} we provide some parameters of our self-consistent fit of the RV and
transit data. This fit relies on only $89$ HQ transits reprocessed in this work. This fit
allows us to estimate possible RV trend in the data (coefficient $c_1$). It is consistent
with zero, and through the Roemer effect it would inspire the following TTV curvature:
\begin{equation}
q^{\rm RA} = (2.6 \pm 3.9)\ \text{Gyr}^{-1}.
\end{equation}
As we can see, this value may only explain a minor portion of the observed $q$. The
uncertainty in the tidal part of $q$ remains almost unchanged as
well.

\section{Discussion}
The nonlinear TTV trend in WASP-4 reveals a confirmed high confidence, even in quite
conservative treatment. Assuming the `separated regularized' noise model, which is more
adaptive, the TTV significance becomes somewhere between $3.8$-sigma (HQ data) and
$4.8$-sigma (full data), depending on how much spots perturb the derived timings.

Therefore, in addition to WASP-12 \citep{Maciejewski18a} this target represents yet another
such example. Still, the trend curvature for WASP-4 is estimated to be only half of what
was claimed initially, and about $6-8$ times more slow than for WASP-12. Based on
formula~(16) from \citep{Bouma19} and our fits, the modified quality factor becomes
$Q_\star' \sim 60000$, with a 1-sigma range of $45000-85000$. Given the large uncertainty,
this is not too much small compared to theoretical predictions. For example, this agrees
well with the estimation $Q_\star' \simeq (1.2{+1.0\atop-0.5})\cdot 10^5$ computed by
\citet{Bouma19} based on \citep{Penev18}. In terms of a more Gaussian parameter
$1/Q_\star'$, the displacement of these two estimations becomes just about $1.2$-sigma.

Simultaneously, there is a wide range of predictions on $Q_\star'$ in the literature, in
particular those yielding a much larger value \citep{CCameron18}. It seems that both the
WASP-12 and WASP-4 cases favour to the models implying rather small $Q_\star'$. WASP-4
transit observations should be continued in order to further refine its TTV.

In this work we considered only a tidal orbit decay model. The TTV interpretation through
tidal apsidal drift is also possible \citep{Patra17}, but it is unlikely to be robustly
distinguished for WASP-4, given the small TTV magnitude.

While this Letter was in review, \citet{Bouma20} reported a radial acceleration of WASP-4
of $-15$~m~s$^{-1}$~yr$^{-1}$, based on new Keck data. This would explain the observed TTV
through the light arrival delay, but they did not use full CORALIE and HARPS data from
\citet{Baluev19}. This full dataset appears strongly inconsistent with their value for
$c_1$, see Fig.~2 in the online-only supplement. The RV trend is only supported by a few
recent Keck observations made after a long pause. This indicates that the RV variation they
captured looks like a severely nonlinear RV variation rather than constant radial
acceleration. This makes their conclusions that RV variation can explain the TTV
disputable. RV was nearly constant at least until 2014, while only recently accumulating
observable bias. Such a nonlinear RV variation may induce a much smaller TTV than constant
acceleration. We believe that results by \citet{Bouma20} need to be reassessed in view of
complete RV data, but this is a work for separate investigation.

\section*{Acknowledgements}
Organization of the EXPANSION project (ENS, IAS), statistical analysis (RVB), and data
collection (VSS), apart from the observations, were supported by Russian Science
Foundation grant 19-72-10023. SH acknowledges CNES grant 837319 support for processing the
observations. Authors would like to thank the TRAPPIST team for sharing their archival
data, as well as the anonymous reviewer for their useful comments and suggestions.

%%%%%%%%%%%%%%%%%%%%%%%%%%%%%%%%%%%%%%%%%%%%%%%%%%



\bibliographystyle{mnras}
\bibliography{wasp4}

%%%%%%%%%%%%%%%%%%%%%%%%%%%%%%%%%%%%%%%%%%%%%%%%%%

%%%%%%%%%%%%%%%%% APPENDICES %%%%%%%%%%%%%%%%%%%%%

%\appendix

%%%%%%%%%%%%%%%%%%%%%%%%%%%%%%%%%%%%%%%%%%%%%%%%%%

% Don't change these lines
\bsp	% typesetting comment
\label{lastpage}
\end{document}